# A new method for phase II monitoring of multivariate simple linear profiles

Seyed Nasser Moosavi

Dr. Mohammad Saleh Owlia

Dr. Ashkan Khalifeh

**Abstract**:

A scope in quality control, which has recently received a great deal of attention is profile that characterizes the quality of a product or process by a relationship between two or more variables. In this paper, we propose an EWMA chart for phase II monitoring of multivariate simple linear profile in which several correlated response variables have linear relationships with one explanatory variable. The statistical performance of this scheme is evaluated in terms of out-of-control average run length. Although it seldom signals for small shifts, it is superior to previous works in detecting moderate and big shifts.

## 1. Introduction:

One of the key factors in success of organizations is the quality of products or services delivered to the customers. Montgomery (2009) stated that the quality is inversely proportional to variability, it means that as variability in characteristics of a product or service dwindles, the quality increases. Developing techniques in statistical process control (SPC) have resulted in more practical and complicated quality control methods. A field in SPC which has recently received a great deal of interest is profile monitoring. When the quality of a product or process can be characterized by a relationship between a response variable and one or more explanatory variables, this relationship is referred to as profile. Among the articles reviewing profiles, we recommend Woodall, Spitzner, Montgomery and Gupta (2004) and Woodall (2007). Also, a review of some existing scopes in SPC, including profile monitoring is given in Woodall and Montgomery (1999, 2014). Some of the applications of profile monitoring in real situations have been explained by Zeng, Neogi and Zhou (2014), Amiri, Jensen and Kazemzadeh (2010), Mahmoud, Parker, Woodall and Hawkins (2007), Noorossana, Eyvazian and Vaghefi (2010), and Kang and Albin (2000).

Although linear profiles are considered as the simplest kind of profile, they are frequently used in industrial and service sectors and the research carried out on this field is really extensive. For example, see Kim, Mahmoud and Woodall (2003), Saghaei, Mehrjoo and Amiri (2009), Magalhaes and Doellinger (2016), Hosseinifard, Abdollahian and Zeephongsekul (2011), Noorossana, Amiri and Soleimani (2008), and Zou, Tsung and Wang (2007). Two approaches were proposed by Kang and Albin (2000). The first approach uses a multivariate $T^2$ to monitor the parameters, slope and intercept, and the latter monitors average residuals, using EWMA and R charts. Kim, Mahmoud and Woodall (2003) proposed a method which is more similar to the second approach of Kang and Albin (2000). Instead of using deviations from the in-control line, they coded the independent variable so that the average value is set to zero, and used the estimated regression coefficients from each sample to construct univariate EWMA charts. Saghaei, Mehrjoo and Amiri (2009) introduced a method based on cumulative sum statistic to monitor linear profiles in phase II. Hosseinifard, Abdollahian and Zeephongsekul (2011) used artificial neural networks to detect and classify the shifts in linear profiles. Noorossana, Amiri and Soleimani (2008) studied a case of linear profiles in which independence assumption is violated, and investigated the effect of autocorrelation between profiles using ARL criterion. Zou, Tsung and Wang (2007) proposed a novel EWMA scheme along with two enhancement features, say, the variable sampling interval and the parametric diagnosis approach to monitor general linear profiles.

Although an extensive research has been carried out into simple linear and nonlinear profiles, monitoring multiple functional profiles in which more than two variables are related together, is yet to be well addressed. Zhang, Ren, Yao, Zou and Wang (2015) proposed a new monitoring scheme incorporating the regression adjustment technique into the functional principal component analysis, to monitor the regression-adjusted residuals of multivariate profiles. Kazemzadeh, Noorossana and Ayoubi (2015) developed a maximum likelihood estimator (MLE) to identify time of linear drifts in the mean of multivariate linear profiles in phase II. Zou, Ning and Tsung (2012) introduced a variable-selection based control scheme to the transformations of estimated parameters in general multivariate linear profiles. This scheme resolves two problems in existing parametric methods, namely low detection ability in the case of large parameter dimensionality and inefficiency in determining the type of parameter change.

Multivariate simple linear profile refers to a kind of profile in which several response variables have relationships with one explanatory variable, and the relationship between each response variable and the explanatory variable is linear. Since all response variables are affected by one explanatory variable, they are correlated and can't be monitored separately. The work here is motivated by a real example at Body Shop of an automotive industry introduced by Noorossana, Eyvazian and Vaghefi (2010). Their purpose was to investigate calibration between desired force and the real forces produced by a hydraulic press machine. This machine has a built-in programmable logic controller (PLC) to adjust input and output factors in order to improve the quality of outputs. Its input is the nominal force that must be exerted by cylinders upon metal plates to give the desired parts. Thereby, for each nominal force (explanatory variable) there exist four real forces (response variables). These forces are collected from four cylinders and measured by PLC. The significant difference between the nominal force and the real forces causes the system to produce defective items. The relationship between the nominal force and the real forces proves to be linear and the real forces are correlated. This is a practical example of multivariate simple linear profiles, where simple refers to one explanatory variable and multivariate refers to several response variables. Ebadi and Amiri (2012) proposed three methods for measuring process capability of multivariate simple linear profiles. Soleimani and Noorossana (2014) studied phase II analysis of multivariate simple linear profiles when the independence assumption of residuals is violated. We have extended the scheme proposed by Huwang, Wang, Xue and Zou (2014) for monitoring the mean of general linear profiles to the case of multivariate simple linear profiles. The rest of the paper is organized as follows; in section 2, multivariate simple linear profile and its basic assumptions are given. In section 3, the proposed monitoring scheme is represented.

Section 4 represents the results of ARL evaluation using simulation, and section 5 concludes the paper.

## 2. Multivariate simple linear profile

In some cases, there are several correlated quality characteristics and all of them are linear functions of one explanatory variable. This kind of profile is referred to as multivariate simple linear profile. Multivariate refers to several correlated response variables and simple refers to one explanatory variable. Totally, there are k samples of observations collected over time and each sample has n observations. In order to simplify the calculations and ARL comparison, we suppose that the explanatory variable has fixed values from sample to sample. In each sample, each of n values of explanatory variable has p corresponding values pertaining to p response variables, hence, the $i^{th}$ ($i = 1, 2, ..., n$) member of the $k^{th}$ sample is $(x_{ik}, y_{i1k}, y_{i2k}, ..., y_{ipk})$. As represented by Noorossana, Eyvazian and Vaghefi (2010), the profile model in the kth sample is as follows:

$$Y_k = X B_k + E_k \tag{1}$$

Where $Y_k = (y_{1k}, y_{2k}, ..., y_{nk})^T$ is a $n \times p$ matrix of response variables, $X = [1, x]$ is a $n \times 2$ matrix of explanatory variable, $B$ is a $2 \times p$ matrix of known parameters (since we are dealing with phase II monitoring, the parameters' values have already been estimated in phase I), and $E_k$ is a $n \times p$ matrix of error terms. The matrix form of equation (1) is as follows:

$$\begin{bmatrix} y_{11k} & y_{12k} & \cdot & \cdot & y_{1pk} \\ y_{21k} & y_{22k} & \cdot & \cdot & y_{2pk} \\ \cdot & \cdot & \cdot & \cdot & \cdot \\ \cdot & \cdot & \cdot & \cdot & \cdot \\ y_{n1k} & y_{n2k} & \cdot & \cdot & y_{npk} \end{bmatrix} = \begin{bmatrix} 1 & x_1 \\ 1 & x_2 \\ \cdot & \cdot \\ \cdot & \cdot \\ 1 & x_n \end{bmatrix} \begin{bmatrix} \beta_{01k} & \beta_{02k} & \cdot & \cdot & \beta_{0pk} \\ \beta_{11k} & \beta_{12k} & \cdot & \cdot & \beta_{1pk} \end{bmatrix} + \begin{bmatrix} \varepsilon_{11k} & \varepsilon_{12k} & \cdot & \cdot & \varepsilon_{1pk} \\ \varepsilon_{21k} & \varepsilon_{22k} & \cdot & \cdot & \varepsilon_{2pk} \\ \cdot & \cdot & \cdot & \cdot & \cdot \\ \cdot & \cdot & \cdot & \cdot & \cdot \\ \varepsilon_{n1k} & \varepsilon_{n2k} & \cdot & \cdot & \varepsilon_{npk} \end{bmatrix} \tag{2}$$

$y_{ijk}$ ($i = 1, 2, ..., n$, $j = 1, 2, ..., p$, $k = 1, 2, ...$) refers to the $i^{th}$ value of the $j^{th}$ response variable in the $k^{th}$ sample. The vector of error terms is assumed to have a multivariate normal distribution with mean vector zero and covariance matrix $\Sigma$ written as follows:

$$\Sigma = \begin{bmatrix} \sigma_{11} & \sigma_{12} & \cdots & \sigma_{1p} \\ \sigma_{21} & \sigma_{22} & \cdots & \sigma_{2p} \\ \vdots & \vdots & \ddots & \vdots \\ \sigma_{p1} & \sigma_{p2} & \cdots & \sigma_{pp} \end{bmatrix} \tag{3}$$

Where $\sigma_{uv}$ is the covariance between $u^{th}$ and $v^{th}$ error terms. Since we are dealing with phase II monitoring schemes, $\Sigma$ and $B_k$ are known or can be estimated from the in-control sample in phase I.

## 3. Control chart for monitoring the mean of the profile

In this section, to monitor the multivariate simple linear profile, we propose an approach monitoring the mean of the profile. Since $E(Y_k|X) = XB_k$ is a $n \times p$ matrix, it's not possible to calculate covariance matrix for that. To obviate this fault, we multiply $E(Y_k|X)_{n \times p}$ by $1_{p \times 1}$ which is a column matrix whose all members are 1. The result is as follows:

$$XB_k 1 = \begin{bmatrix} 1 & x_1 \\ 1 & x_2 \\ . & . \\ . & . \\ 1 & x_n \end{bmatrix} \begin{bmatrix} \beta_{01k} & \beta_{02k} & . & . & \beta_{0pk} \\ \beta_{11k} & \beta_{12k} & . & . & \beta_{1pk} \end{bmatrix} \begin{bmatrix} 1 \\ 1 \\ . \\ . \\ 1 \end{bmatrix}$$

$$= \begin{bmatrix} \beta_{01k}+\beta_{11k}x_1 & \beta_{02k}+\beta_{12k}x_1 & . & . & \beta_{0pk}+\beta_{1pk}x_1 \\ \beta_{01k}+\beta_{11k}x_2 & \beta_{02k}+\beta_{12k}x_2 & . & . & \beta_{0pk}+\beta_{1pk}x_2 \\ . & . & . & . & . \\ . & . & . & . & . \\ \beta_{01k}+\beta_{11k}x_n & \beta_{02k}+\beta_{12k}x_n & . & . & \beta_{0pk}+\beta_{1pk}x_n \end{bmatrix} \begin{bmatrix} 1 \\ 1 \\ . \\ . \\ 1 \end{bmatrix}$$

$$= \begin{bmatrix} \beta_{01k}+\beta_{11k}x_1+\beta_{02k}+\beta_{12k}x_1+...+\beta_{0pk}+\beta_{1pk}x_1 \\ \beta_{01k}+\beta_{11k}x_2+\beta_{02k}+\beta_{12k}x_2+...+\beta_{0pk}+\beta_{1pk}x_2 \\ . \\ . \\ \beta_{01k}+\beta_{11k}x_n+\beta_{02k}+\beta_{12k}x_n+...+\beta_{0pk}+\beta_{1pk}x_n \end{bmatrix} \quad (4)$$

Suppose that $X_i = \begin{bmatrix} 1 & x_i \end{bmatrix}$ is the $i^{th}$ row of the matrix $X$. Consequently, $X_i \hat{B}_k 1$ is the $i^{th}$ row of the matrix $X \hat{B}_k 1$ and represents sum of the means of all p response variables in the $i^{th}$ observation of the $k^{th}$ sample. Now, to calculate $Var(X_i \hat{B}_k 1)$, we perform as follows:

$$Var(X_i \hat{B}_k 1) = X_i Var(\hat{B}_k 1) X_i^T = X_i \Sigma_B X_i^T \quad (5)$$

and

$$\Sigma_B = Var(\hat{B}_k 1) = Var \begin{bmatrix} \hat{\beta}_{01k} + \hat{\beta}_{02k} + ... + \hat{\beta}_{0pk} \\ \hat{\beta}_{11k} + \hat{\beta}_{12k} + ... + \hat{\beta}_{1pk} \end{bmatrix} = \begin{bmatrix} \sigma^B_{11} & \sigma^B_{12} \\ \sigma^B_{21} & \sigma^B_{22} \end{bmatrix} \quad (6)$$

where

$$\sigma^B_{11} = Var(\hat{\beta}_{01k} + \hat{\beta}_{02k} + ... + \hat{\beta}_{0pk}) = (\frac{1}{n} + \frac{\overline{x}^2}{S_{xx}}) \left[ \sum_{\substack{u=1 \\ u=v}}^{p} \sum_{v=1}^{p} \sigma_{uv} + 2 \sum_{\substack{u=1 \\ u<v}}^{p} \sum_{v=1}^{p} \sigma_{uv} \right] \quad (7)$$

$$\sigma^B_{22} = Var(\hat{\beta}_{11k} + \hat{\beta}_{12k} + ... + \hat{\beta}_{1pk}) = (\frac{1}{S_{xx}}) \left[ \sum_{\substack{u=1 \\ u=v}}^{p} \sum_{v=1}^{p} \sigma_{uv} + 2 \sum_{\substack{u=1 \\ u<v}}^{p} \sum_{v=1}^{p} \sigma_{uv} \right] \quad (8)$$

$$\sigma^B_{12} = Cov(\hat{\beta}_{01k} + \hat{\beta}_{02k} + ... + \hat{\beta}_{0pk}, \hat{\beta}_{11k} + \hat{\beta}_{12k} + ... + \hat{\beta}_{1pk}) = -\frac{\overline{x}}{S_{xx}} \left[ \sum_{\substack{u=1 \\ u=v}}^{p} \sum_{v=1}^{p} \sigma_{uv} + 2 \sum_{\substack{u=1 \\ u<v}}^{p} \sum_{v=1}^{p} \sigma_{uv} \right] \quad (9)$$

$\sigma_{uv}$ is the $u^{th}$ row of the $v^{th}$ column of covariance matrix $\Sigma$. The proof is given in appendix.

So, we have $Var(X_i \hat{B}_k 1) = X_i \Sigma_B X_i^T$ and $E(X_i \hat{B}_k 1) = X_i B_k 1$. When the process is in control:

$$X_i \hat{B}_k 1 \sim N(X_i B_k 1, X_i \Sigma_B X_i^T) \quad (10)$$

Consequently, for a given $\alpha$, there exists a $m_\alpha$ so that:

$$\Pr\{\max | \frac{X_i \hat{B}_k 1 - X_i B_k 1}{\sqrt{X_i \Sigma_B X_i^T}} | > m_\alpha \} = \alpha \quad (11)$$

or

$$\Pr\{X_i \hat{B}_k 1 \in X_i B_k 1 \pm m_\alpha \sqrt{X_i \Sigma_B X_i^T} \} = 1 - \alpha \quad (12)$$

Equations (11) and (12) are derived from similar ones in Huwang, Wang, Xue and Zou (2014) who monitored general linear profiles using simultaneous confidence sets. Now we can use the confidence set, $(X_i B_k 1 \pm m_\alpha \sqrt{X_i \Sigma_B X_i^T})$ to monitor the mean of the profile. If $X_i B_k 1$ is not completely included in the confidence set, the corresponding chart triggers an out-of-control signal. A merit of this method is monitoring sum of the means of all p response variables together and if one of these response variables is out of control, the chart triggers a signal for the entire profile. In order to increase the sensitivity of the control chart, we define a control chart based on EWMA as follows:

$$\left(\hat{B}_k 1\right)_e (j) = \theta \left(\hat{B}_k 1\right)_j + (1-\theta)\left(\hat{B}_k 1\right)_e (j-1) \tag{13}$$

Where $\left(\hat{B}_k 1\right)_e (0)$ is the in-control value $B_k 1$. When the process is in control, it's easy to show that:

$$\left(\hat{B}_k 1\right)_e (j) \sim N\left(B_k 1, \frac{\theta}{2-\theta}\left(1-(1-\theta)^{2j}\right)\Sigma_B\right) \tag{14}$$

Define

$$V(j) = max \left| \frac{X_i \left(\hat{B}_k 1\right)_e (j) - X_i B_K 1}{\sqrt{X_i \Sigma_B X_i^T}} \right| \tag{15}$$

Hence, the control chart for $XB_k 1$ triggers an out-of-control signal if

$$V(j) > L_B \sqrt{\frac{\theta}{2-\theta}\left(1-(1-\theta)^{2j}\right)} \tag{16}$$

Where $L_B$ is gained from a known predetermined ARL.

## 4. ARL comparison

In this section, the performance of proposed monitoring scheme is assessed using simulation. The common criterion for assessing the performance of different schemes is out-of-control (OC) ARL representing the mean number of the samples taken to recognize OC condition after a shift occurs. Considering $L_B = 3.6233$, we have in-control (IC) ARL=200. The smoothing constant $\theta$ is set equal to 0.2. ARL performance of the proposed scheme will be evaluated using 5000 replications. To keep the pace with previous works, we consider the case in which $p = 2$ and the underlying model is as follows:

$$y_1 = 3 + 2x + \varepsilon_1$$
$$y_2 = 2 + 1x + \varepsilon_2 \quad (17)$$

Where $x_i$-values are equal to 2,4,6 and 8. The error terms vector $(\varepsilon_1, \varepsilon_2)$ is a bivariate normal random vector whose mean vector is zero and covariance matrix is $\Sigma = \begin{bmatrix} \sigma_1^2 & \rho\sigma_1\sigma_2 \\ \rho\sigma_1\sigma_2 & \sigma_2^2 \end{bmatrix}$. To facilitate comparison with previous works, we consider $\sigma_1^2 = 1$, $\sigma_2^2 = 1$ and $\rho = 0.1, 0.5, 0.9$.

Different types of shifts are considered to evaluate the performance of proposed monitoring scheme. To facilitate comparison with the schemes proposed by Noorossana, Eyvazian and Vaghefi (2010), we have given the out-of-control ARL values of their first scheme in parentheses. Table 1 represents ARL values of the monitoring scheme for different shifts in $\beta_{01}$ in units of $\sigma_1$. In comparison to the scheme of Noorossana, Eyvazian and Vaghefi (2010), when $\lambda \leq 1$, this chart is not efficient enough, but as $\lambda$ increases, its performance improves. when $\lambda > 1$ and $\rho = 0.1$, the proposed scheme is superior, specially for $\lambda \geq 1.4$ where it results in ARL=1. By contrast with the methods proposed by Noorossana, Eyvazian and Vaghefi (2010), as $\rho$ value increases, the chart's power in detecting shifts declines.

**Table 1**

The simulated ARL values when $\beta_{01}$ shifts to $\beta_{01} + \lambda\sigma_1$ (in-control ARL=200).

| $\lambda$ | 0.2 | 0.4 | 0.6 | 0.8 | 1 | 1.2 | 1.4 | 1.6 | 1.8 | 2 |
|---|---|---|---|---|---|---|---|---|---|---|
| $\rho = 0.1$ | 199.57 (66.6) | 198.06 (19.17) | 191.21 (9.2) | 127.63 (5.9) | 9.90 (4.3) | 1.21 (3.5) | 1.00 (2.9) | 1.00 (2.5) | 1.00 (2.3) | 1.00 (2.0) |
| $\rho = 0.5$ | 199.61 (53.9) | 198.64 (14.4) | 195.40 (7.3) | 180.54 (4.9) | 80.66 (3.7) | 6.64 (3.0) | 1.07 (2.5) | 1.00 (2.2) | 1.00 (2.0) | 1.00 (1.9) |
| $\rho = 0.9$ | 199.76 (14.8) | 198.93 (4.9) | 197.01 (3.0) | 190.67 (2.2) | 154.28 (1.9) | 39.75 (1.6) | 3.54 (1.3) | 1.03 (1.0) | 1.00 (1.0) | 1.00 (1.0) |

Table 2 represents ARL values for different shifts in $\beta_{02}$. Like the previous case, when $\lambda > 1$, the chart's ability in detecting shifts profoundly improves and positive change in $\rho$ value decreases the chart's sensitivity.

**Table 2**

The simulated ARL values when $\beta_{02}$ shifts to $\beta_{02} + \lambda\sigma_1$ (in-control ARL=200).

| $\lambda$ | 0.2 | 0.4 | 0.6 | 0.8 | 1 | 1.2 | 1.4 | 1.6 | 1.8 | 2 |
|---|---|---|---|---|---|---|---|---|---|---|
| $\rho = 0.1$ | 199.54 | 198.27 | 191.16 | 129.31 | 10.50 | 1.09 | 1.00 | 1.00 | 1.00 | 1.00 |
| $\rho = 0.5$ | 199.82 | 198.78 | 195.60 | 180.66 | 78.91 | 6.34 | 1.28 | 1.00 | 1.00 | 1.00 |
| $\rho = 0.9$ | 199.55 | 199.01 | 197.13 | 190.52 | 154.56 | 38.71 | 3.36 | 1.01 | 1 | 1.00 |

Table 3 indicates ARL values for shifts in $\beta_{11}$. When $\lambda = 0.25$, the chart works best. Considering $\rho = 0.1$, when $\lambda \leq 0.175$ the current chart is less efficient than that proposed by Noorossana, Eyvazian and Vaghefi (2010), however when $\lambda \geq 0.2$, it outperforms it.

**Table 3**

The simulated ARL values when $\beta_{11}$ shifts to $\beta_{11} + \lambda\sigma_1$ (in-control ARL=200).

| $\lambda$ | 0.025 | 0.050 | 0.075 | 0.1 | 0.125 | 0.15 | 0.175 | 0.2 | 0.225 | 0.25 |
|---|---|---|---|---|---|---|---|---|---|---|
| $\rho = 0.1$ | 199.87 (108.5) | 199.25 (39.4) | 198.00 (17.7) | 194.67 (10.6) | 182.22 (7.4) | 114.82 (5.6) | 21.28 (4.6) | 2.44 (3.8) | 1.03 (3.4) | 1.00 (3.0) |
| $\rho = 0.5$ | 199.92 (91.0) | 199.55 (30.1) | 198.48 (13.9) | 197.16 (8.6) | 192.62 (6.1) | 178.76 (4.7) | 116.16 (3.9) | 28.77 (3.3) | 4.29 (2.9) | 1.35 (2.6) |
| $\rho = 0.9$ | 200.11 (30.2) | 199.57 (8.6) | 198.76 (4.8) | 197.75 (3.3) | 195.78 (2.6) | 189.96 (2.2) | 171.01 (2.0) | 105.33 (1.8) | 29.86 (1.6) | 5.80 (1.4) |

Tables 4, 5 and 6 represent ARL values, when two of the parameters have simultaneously changed. Similar to previous tables, as $\rho$ value increases, the chart's efficiency decreases.

**Table 4**

The simulated ARL values under combination of intercepts shifts from $\beta_{01}$ to $\beta_{01} + \lambda_1\sigma_1$ and $\beta_{02}$ to $\beta_{02} + \lambda_2\sigma_1$ (in-control ARL=200)

| $\lambda_1$ \ $\lambda_2$ | 0.3 | 0.35 | 0.4 | 0.45 | 0.5 | 0.55 | 0.6 |
|---|---|---|---|---|---|---|---|
| 0.3 | 191.25 195.50 197.23 (25.2) | 186.11 193.67 196.10 | 177.10 191.54 194.73 (17.8) | 159.71 187.17 193.18 | 128.47 180.20 190.52 (12.6) | 84.91 168.75 186.37 | 45.84 148.55 180.87 |
| 0.35 | 186.26 193.84 196.20 | 176.93 191.26 195.00 | 160.05 187.24 192.71 | 127.95 180.30 190.77 | 84.84 169.32 186.48 | 45.78 148.81 180.96 | 22.21 117.26 170.47 |
| 0.4 | 177.07 191.53 194.58 (17.8) | 159.55 187.45 192.95 | 128.66 180.50 190.59 (14.4) | 84.78 168.88 186.43 | 48.09 148.76 180.35 (11.4) | 22.66 116.79 171.08 | 10.52 80.54 153.47 |
| 0.45 | 159.12 187.46 193.08 | 128.30 180.20 190.63 | 86.01 168.18 187.06 | 47.37 147.88 179.96 | 22.13 117.51 169.65 | 10.41 81.18 154.39 | 3.93 48.35 127.34 |
| 0.5 | 127.19 179.94 190.29 (12.4) | 86.94 168.48 186.39 | 48.59 147.67 180.44 (11.5) | 22.45 116.10 170.11 | 10.42 80.98 153.95 (9.9) | 4.69 48.39 128.47 | 2.20 25.10 95.79 |
| 0.55 | 86.66 168.25 | 47.47 149.09 | 23.51 117.12 | 10.91 80.54 | 4.26 48.31 | 1.98 25.59 | 1.43 12.26 |
|     | 186.76 | 180.64 | 170.52 | 154.44 | 128.98 | 95.46  | 63.64 |
| 0.6 | 46.39  | 22.98  | 9.52   | 4.30   | 2.30   | 1.24   | 1.13  |
|     | 147.21 | 116.98 | 79.79  | 47.88  | 25.57  | 12.68  | 6.66  |
|     | 180.67 | 170.45 | 153.79 | 129.32 | 96.41  | 65.37  | 39.66 |

**Table 5**

The simulated ARL values under combination of slopes shifts from $\beta_{11}$ to $\beta_{11} + \lambda_1 \sigma_1$ and $\beta_{12}$ to $\beta_{12} + \lambda_2 \sigma_1$ (in-control ARL=200)

| $\lambda_1$ \ $\lambda_2$ | 0.02 | 0.04 | 0.06 | 0.08 | 0.09 | 0.1 |
|---|---|---|---|---|---|---|
| 0.02 | 199.43 | 198.78 | 197.50 | 194.45 | 191.64 | 186.64 |
|      | 199.55 | 199.29 | 198.46 | 197.22 | 195.86 | 193.68 |
|      | 199.62 | 199.48 | 198.82 | 198.06 | 197.02 | 196.37 |
|      | (116.6) | (58.6) | (26.9) | (14.9) |        | (9.8) |
| 0.04 | 198.94 | 197.42 | 194.82 | 186.63 | 176.63 | 154.07 |
|      | 199.36 | 198.26 | 196.82 | 194.01 | 191.45 | 186.79 |
|      | 199.34 | 198.76 | 197.92 | 196.44 | 194.97 | 192.86 |
|      | (58.5) | (45.3) | (27.0) | (16.1) |        | (10.6) |
| 0.06 | 197.61 | 194.87 | 186.87 | 155.08 | 115.97 | 68.26 |
|      | 198.49 | 196.92 | 193.90 | 186.44 | 178.52 | 163.28 |
|      | 198.74 | 197.90 | 196.45 | 193.13 | 189.98 | 185.11 |
|      | (27.1) | (26.9) | (21.4) | (14.7) |        | (10.6) |
| 0.08 | 194.79 | 186.73 | 154.61 | 68.60  | 34.10  | 14.20 |
|      | 197.16 | 194.17 | 186.66 | 162.14 | 135.80 | 96.25 |
|      | 198.04 | 196.35 | 192.91 | 184.97 | 176.95 | 162.95 |
|      | (15.0) | (16.3) | (15.0) | (12.6) |        | (9.8) |
| 0.09 | 191.80 | 176.43 | 114.95 | 33.09  | 13.90  | 5.84  |
|      | 195.88 | 191.28 | 178.04 | 135.93 | 98.39  | 57.99 |
|      | 197.30 | 194.90 | 190.08 | 177.04 | 162.40 | 138.38 |
| 0.1  | 186.62 | 154.37 | 68.13  | 13.95  | 5.88   | 2.47  |
|      | 194.19 | 186.81 | 163.50 | 97.14  | 56.97  | 28.75 |
|      | 196.19 | 192.94 | 185.24 | 162.34 | 140.17 | 105.87 |
|      | (9.9)  | (10.5) | (10.6) | (9.9)  |        | (8.5) |

**Table 6**

The simulated ARL values under combination of slope and intercept shifts from $\beta_{01}$ to $\beta_{01} + \lambda_1 \sigma_1$ and $\beta_{11}$ to $\beta_{11} + \lambda_2 \sigma_1$ (in-control ARL=200)

| $\lambda_1$ \ $\lambda_2$ | 0.02 | 0.04 | 0.06 | 0.08 | 0.1 | 0.12 | 0.14 |
|---|---|---|---|---|---|---|---|
| 0.1 | 199.55 | 199.06 | 197.67 | 195.43 | 188.32 | 161.87 | 81.73 |
|     | 199.73 | 199.16 | 198.52 | 197.20 | 194.44 | 187.69 | 167.96 |
|     | 199.72 | 199.57 | 198.90 | 197.99 | 196.38 | 193.82 | 186.43 |
|     | (51.4) | (23.3) | (13.2) | (8.9)  | (6.7)  |        |        |
| 0.2 | 199.00 | 197.94 | 195.45 | 189.38 | 167.18 | 93.85  | 22.35 |
|     | 199.10 | 198.72 | 197.23 | 194.85 | 189.08 | 172.55 | 116.93 |
|     | 199.21 | 198.87 | 198.07 | 196.74 | 194.13 | 187.73 | 170.18 |
|     | (24.6) | (13.8) | (9.3)  | (6.9)  | (5.5)  |        |        |
| 0.3 | 197.86 | 195.63 | 189.91 | 171.23 | 103.96 | 27.66  | 4.00  |
|     | 198.59 | 197.34 | 195.23 | 189.91 | 174.86 | 125.06 | 50.93 |
|     | 198.95 | 198.26 | 196.99 | 194.38 | 188.56 | 173.73 | 132.28 |
|     | (14.6) | (9.6)  | (7.1)  | (5.6)  | (4.6)  |        |        |
| 0.4 | 195.92 | 190.62 | 174.07 | 111.22 | 30.74 | 5.90 | 1.63 |
| | 197.57 | 195.36 | 190.47 | 176.77 | 131.95 | 56.02 | 13.31 |
| | 198.46 | 196.88 | 194.44 | 189.58 | 175.85 | 139.58 | 68.75 |
| | (9.9) | (7.2) | (5.7) | (4.7) | (4.0) | | |
| 0.5 | 191.10 | 175.61 | 119.47 | 38.05 | 6.83 | 1.63 | 1.04 |
| | 195.31 | 190.78 | 178.27 | 138.24 | 63.79 | 16.28 | 3.59 |
| | 197.09 | 194.76 | 189.60 | 177.94 | 143.12 | 75.19 | 24.88 |
| | (7.3) | (5.8) | (4.8) | (4.1) | (3.6) | | |

Tables 7 and 8 indicate the results of simulation for shifts in $\sigma_1$ and both $\sigma_1$ and $\sigma_2$, respectively.

**Table 7**

The simulated ARL values when $\sigma_1$ shifts to $\lambda\sigma_1$ (in-control ARL=200)

| $\lambda$ | 1.2 | 1.4 | 1.6 | 1.8 | 2.0 | 2.2 | 2.4 | 2.6 | 2.8 | 3.0 |
|---|---|---|---|---|---|---|---|---|---|---|
| $\rho = 0.1$ | 197.31 | 156.89 | 112.36 | 52.31 | 12.34 | 1.36 | 1.13 | 1.00 | 1.00 | 1.00 |
| $\rho = 0.5$ | 198.30 | 160.31 | 112.31 | 54.03 | 13.61 | 2.31 | 1.26 | 1.12 | 1.00 | 1.00 |
| $\rho = 0.9$ | 199.30 | 165.20 | 114.96 | 57.30 | 14.31 | 2.75 | 1.65 | 1.36 | 1.10 | 1.00 |

**Table 8**

The simulated ARL values under combination of standard deviations' shifts from $\sigma_1$ to $\lambda_1\sigma_1$ and from $\sigma_2$ to $\lambda_2\sigma_2$ (in-control ARL=200)

| $\lambda_1$ \ $\lambda_2$ | 1.1 | 1.2 | 1.3 | 1.4 | 1.5 |
|---|---|---|---|---|---|
| 1.1 | 141.42 | 68.12 | 33.21 | 9.36 | 3.21 |
| | 152.31 | 74.87 | 38.26 | 12.32 | 4.59 |
| | 167.65 | 80.31 | 43.91 | 15.85 | 6.89 |
| | (81.6) | (55.5) | (39.1) | (29.0) | (22.0) |
| 1.2 | 68.26 | 33.96 | 9.39 | 3.46 | 2.56 |
| | 74.78 | 38.21 | 12.38 | 4.56 | 3.56 |
| | 80.30 | 44.31 | 15.78 | 6.89 | 5.61 |
| | (55.6) | (41.9) | (31.9) | (24.3) | (19.4) |
| 1.3 | 33.23 | 9.32 | 3.31 | 2.59 | 1.14 |
| | 38.13 | 12.36 | 4.46 | 3.26 | 1.88 |
| | 44.02 | 15.23 | 6.90 | 5.46 | 3.13 |
| | (39.7) | (32.5) | (25.3) | (21.0) | (17.2) |
| 1.4 | 9.48 | 3.29 | 2.65 | 1.13 | 1.08 |
| | 12.45 | 4.53 | 3.12 | 1.89 | 1.61 |
| | 15.26 | 6.82 | 5.56 | 3.12 | 2.56 |
| | (29.2) | (24.3) | (20.4) | (17.4) | (14.8) |
| 1.5 | 3.3364 | 2.5614 | 1.1460 | 1.0923 | 1.0112 |
| | 4.4531 | 3.1278 | 1.8813 | 1.5813 | 1.3643 |
| | 6.7826 | 5.5316 | 3.1157 | 2.5124 | 1.4516 |
| | (22.2) | (19.5) | (17.1) | (15.3) | (13.0) |

## 5. Conclusion

In this paper, a control chart scheme for phase II monitoring of multivariate simple linear profiles is proposed. This method is an expansion of the method proposed by Huwang, Wang, Xue and Zou (2014) for monitoring general linear profiles. Simulation studies prove that this scheme is extremely sensitive to rather big shifts, in comparison to previous works, however other methods outperform it when the shifts are small. Moreover, it is shown that as $\rho$ value (correlation between error terms) increases, the chart's efficiency decreases.

## Appendix

$\hat{\beta}_{0uk}$ and $\hat{\beta}_{0vk}$ are the least square estimators for the intercepts of $u^{th}$ and $v^{th}$ profiles, respectively. Their covariance is defined as follows:

$$\text{cov}(\hat{\beta}_{0uk}, \hat{\beta}_{0vk}) = \text{cov}(\bar{y}_{.uk} - \hat{\beta}_{1uk}\bar{x}, \bar{y}_{.vk} - \hat{\beta}_{1vk}\bar{x})$$

$$= \text{cov}\left(\sum_{i=1}^{n}\left[\frac{1}{n} + \frac{(x_i - \bar{x})\bar{x}}{S_{xx}}\right] y_{iuk}, \sum_{i=1}^{n}\left[\frac{1}{n} + \frac{(x_i - \bar{x})\bar{x}}{S_{xx}}\right] y_{ivk}\right)$$

$$= \sum_{i=1}^{n}(\frac{1}{n} + \frac{(x_i - \bar{x})\bar{x}}{S_{xx}})^2 \text{cov}(y_{iuk}, y_{ivk}) = \sigma_{uv}\sum_{i=1}^{n}\left(\frac{1}{n} + \frac{(x_i - \bar{x})\bar{x}}{S_{xx}}\right)^2$$

$$= \sigma_{uv}(\frac{1}{n} + \frac{\bar{x}^2}{S_{xx}}) = \rho_{uv}\sigma_u\sigma_v(\frac{1}{n} + \frac{\bar{x}^2}{S_{xx}}).$$

$\hat{\beta}_{1uk}$ and $\hat{\beta}_{1vk}$ are the least square estimators for the slopes of $u^{th}$ and $v^{th}$ profiles, respectively. Their covariance is defined as follows:

$$\text{cov}(\hat{\beta}_{1uk}, \hat{\beta}_{1vk}) = \text{cov}\left(\frac{S_{xy(u)}}{S_{xx}}, \frac{S_{xy(v)}}{S_{xx}}\right) = \frac{1}{(S_{xx})^2}\text{cov}\left(\sum_{i=1}^{n}(x_i - \bar{x})y_{iuk}, \sum_{i=1}^{n}(x_i - \bar{x})y_{ivk}\right)$$

$$= \frac{\sum_{i=1}^{n}(x_i - \bar{x})^2 \text{cov}(y_{iuk}, y_{ivk})}{(S_{xx})^2} = \frac{\sigma_{uv}\sum_{i=1}^{n}(x_i - \bar{x})^2}{(S_{xx})^2} = \frac{\sigma_{uv}}{S_{xx}} = \frac{\rho_{uv}\sigma_u\sigma_v}{S_{xx}}.$$

$\hat{\beta}_{0uk}$ and $\hat{\beta}_{1vk}$ are the least square estimators for the intercept of $u^{th}$ profile and slope of $v^{th}$ profiles, respectively. Their covariance is defined as follows:

$$\text{cov}(\hat{\beta}_{0uk}, \hat{\beta}_{1vk}) = \text{cov}(\bar{y}_{.uk} - \hat{\beta}_{1uk}\bar{x}, \hat{\beta}_{1vk}) = \text{cov}(\bar{y}_{.uk}, \hat{\beta}_{1vk}) - \sigma_{uv}\frac{\bar{x}}{s_{xx}}$$

$$= \text{cov}\left(\sum_{i=1}^{n}\frac{1}{n}y_{iuk}, \sum_{i=1}^{n}\frac{1}{s_{xx}}(x_i - \bar{x})y_{ivk}\right) - \sigma_{uv}\frac{\bar{x}}{s_{xx}} = \frac{\sigma_{uv}\sum_{i=1}^{n}(x_i - \bar{x})}{n(s_{xx})} - \sigma_{uv}\frac{\bar{x}}{s_{xx}}$$

$$= 0 - \sigma_{uv}\frac{\bar{x}}{s_{xx}} = -\sigma_{uv}\frac{\bar{x}}{s_{xx}}.$$

Using the above equations, it's easy to calculate equations (7), (8) and (9), and consequently matrix $\Sigma_B$.